\documentclass[aps,showkeys,nofootinbib]{revtex4}
\RequirePackage[colorlinks,hyperindex]{hyperref}
\RequirePackage[english]{babel}
\RequirePackage[latin1]{inputenc}
\RequirePackage[T1]{fontenc}
\usepackage{graphicx}
\RequirePackage{mathrsfs}
\RequirePackage{amsmath}
\RequirePackage{amssymb}
\RequirePackage{amsbsy}
\RequirePackage{color}
\RequirePackage{bm}
\newcommand{\be}{\begin{equation}}
\newcommand{\ee}{\end{equation}}
\newcommand{\bea}{\begin{eqnarray}}
\newcommand{\eea}{\end{eqnarray}}
\def\de#1/de#2{\frac{\partial {#1}}{\partial {#2}}}
\hypersetup{colorlinks=true,breaklinks=true,urlcolor=blue,linkcolor=red}
\begin{document}
\title{Integrability of Dirac equations in static spherical space-times}
\author{Roberto Cianci\footnote{E-mail: roberto.cianci@unige.it},
Stefano Vignolo\footnote{E-mail: stefano.vignolo@unige.it},
Luca Fabbri\footnote{E-mail: luca.fabbri@edu.unige.it}}
\affiliation{DIME, Universit\`{a} di Genova, Via all'Opera Pia 15, 16145 Genova, ITALY\\
INFN, Sezione di Genova, Via Dodecaneso 33, 16146 Genova, ITALY}
\date{\today}
\begin{abstract}
We consider the Dirac equations in static spherically-symmetric space-times, and we present a type of spinor field whose structure allows the separation of elevation angle and radial coordinate in very general situations. We demonstrate that after such a separation of variables the Dirac equations reduce to two equations that can always be integrated, at least in principle. To prove that ours is a fully-working method, we find an explicit exact solution in the special case of the de Sitter universe.
\end{abstract}
\maketitle
\section{Introduction}
The Dirac theory is arguably one of the most far-reaching physical theory, both for its role in the standard model of high-energy particles and for modelling low-energy systems \cite{hasan2010colloquium, boada2014dirac, suchet2016analog, novoselov2005two, katsnelson2006chiral, schwerdtfeger2015relativistic, pavsteka2017relativistic, autschbach2012perspective}. Yet, finding exact solutions to the Dirac equations is still a rather complicated problem, at least in general $(3+1)$-dimensional space-times.

One feature that seems to be particularly problematic is the fact that the Dirac spinor, being characterized by an intrinsic spin axial-vector, naturally selects a specific spatial direction, so that the consequent lack of isotropy seems to be compatible only with axially-symmetric metrics \cite{Cianci:2019oor, Cianci:2016pvd, Cianci:2015pba}. Just the same, the quest for exact solutions with spherical symmetry retains a certain appeal due to the simplicity of such material distributions.

Correspondingly, in the past there have been several attempts at finding spinor fields that are compatible with a spherically-symmetric background. Because the presence of the spin axial-vector breaks the rotational invariance of the spinor system, any attempt at finding spinor fields that are compatible with spherical symmetry must limit the impact of the spin structures on the background. There are two ways in which this can be done, the first consisting in the inclusion of self-interactions of specific structure that can cancel all spin contributions \cite{Soler:1970xp, Ranada:1983zqr, Saha:2018ufp, Bronnikov:2019nqa}. The second consists in considering multi-particle situations so that the spins cancel, either by averaging out in random configurations \cite{Herdeiro:2017fhv,Herdeiro:2019mbz, Dzhunushaliev:2019kiy, Blazquez-Salcedo:2019qrz, bs-k} or because two spinors are taken as a singlet state where the spins are opposite \cite{fss, fss1, fss2, fss3}.

Recently, an interesting application to Dirac particles in materials having non-trivial curvature was discussed \cite{Bagchi:2023cpz}.

In the special de Sitter case, exact solutions are discussed for example in \cite{Dirac, H, KT, Cotaescu:2007xv, Kanno:2016qcc, Dappiaggi:2019oew}. The approach followed in these references, however, is heavily dependent on the fact that the underlying metric be de Sitter in strict sense.

In the present work we wish to present a procedure that would ensure the Dirac spinor field equation to be integrated in static and spherically-symmetric space-times, without any further assumptions whether they are on interactions or on spin configurations. By extrapolating the general structure of the spinor field that is compatible with Schwarzschild-like metrics, we will see that the corresponding Dirac equations will split, reducing to an angular part that can always be integrated, and a radial part that can always be converted to a Riccati equation, and so always integrable.

We show the viability of our method with a specific example, given by the de Sitter case. The fact that in the specific de Sitter case our method gives results that are similar to those obtained in the literature under more restrictive conditions is to be seen as a sign of consistency of our approach.
\section{Dirac equations with static spherical symmetry}
The initial step of our construction is that of writing the Dirac spinor field equations in static and spherically-symmetric space-time metrics. In spherical coordinates these are given by
\begin{equation}\label{Lem}
g= - e^{2B}\,dr\otimes dr - r^2\,d\theta\otimes d\theta - r^2\sin^2\theta\,d\varphi\otimes d\varphi + e^{2A}\, dt\otimes dt
\end{equation}
where $B(r)$ and $A(r)$ are functions of the radial variable $r$ alone. A basic argument on the Killing structure of this space-time suggests that the temporal/azimuthal dependence be split from the elevation/radial dependence, but we still retain the right to have some freedom in choosing the frame in the $(t-\varphi)$ and the $(\theta-r)$ plane. Consequently we pick the dual sets of frame and coframe given by
\begin{eqnarray}\label{defcoframe}
&&e^1:= \sin\rho(r,\theta)e^B\,dr - r\cos\rho(r,\theta)\,d\theta, \quad e^2:= \sinh\alpha(r,\theta)e^A\,dt + r\sin\theta\cosh\alpha(r,\theta)\,d\varphi \nonumber\\ 
&&e^3:= -\cos\rho(r,\theta)e^B\,dr -r\sin\rho(r,\theta)\,d\theta, \quad e^4:= \cosh\alpha(r,\theta)e^A\,dt + r\sin\theta\sinh\alpha(r,\theta),d\varphi
\end{eqnarray}
and
\begin{eqnarray}\label{defframe}
&&e_1 := \sin\rho(r,\theta)e^{-B}\,\de /de{r} - \frac{\cos\rho(r,\theta)}{r}\,\de /de{\theta}, \quad e_2:= \frac{\cosh\alpha(r,\theta)}{r\sin\theta}\,\de /de{\varphi} - \sinh\alpha(r,\theta)e^{-A}\,\de /de{t} \nonumber\\
&&e_3:= - \cos\rho(r,\theta)e^{-B}\,\de /de{r} - \frac{\sin\rho(r,\theta)}{r}\,\de /de{\theta}, \quad e_4:= - \frac{\sinh\alpha(r,\theta)}{r\sin\theta}\,\de /de{\varphi} + \cosh\alpha(r,\theta)e^{-A}\,\de /de{t}
\end{eqnarray}
where $\rho(r,\theta)$ and $\alpha(r,\theta)$ are suitable functions of the coordinates $r$ and $\theta$ only. Of course $\eta_{\mu\nu}e^\mu\otimes e^\nu =g$ with $\mu,\nu =1,\ldots,4$ and $\eta_{\mu\nu}={\rm diag}(-1,-1,-1,1)$ setting our convention on the signature of the Minkowski metric.

We recall that, given a space-time $M$ endowed with a metric tensor $g_{ij}$ and a linear connection $\Gamma_{ij}^{\,\,\,h}$, the covariant derivative of a spinor field $\psi$ is defined as
\begin{equation}\label{derspin}
D_i\psi = \de{\psi}/de{x^i} - \Omega_i\psi
\end{equation}
where
\begin{equation}\label{spinconnection}
\Omega_i = -\frac{1}{4}g_{jh}\left(\Gamma_{ik}^{\,\,\,j} - e^j_\mu\de{e^\mu_k}/de{x^i}\right)Y^hY^k
\end{equation}
are the coefficients of the spin derivative associated with the metric tensor $g_{ij}$ and a linear connection $\Gamma_{ij}^{\,\,\,h}$ with $Y^i:=\gamma^\mu e_\mu^i$ (in standard notation we would have $Y^i:=\gamma^i$ but we prefer to keep these two matrices distinguished for clarity). The Clifford matrices $\gamma^\mu$ are of course those belonging to the Clifford algebra (see \cite{FMB} for general identities holding among them). For brevity, hereafter we adopt the notation $f_x:=\partial{f}/\partial{x}$ to indicate the partial derivative of the function $f$ with respect to the $x$ variable. Then, the metric \eqref{Lem}, its Levi--Civita connection and the tetrad fields \eqref{defcoframe} and \eqref{defframe} induce coefficients of the spinorial derivative expressed as
\begin{subequations}\label{DOmega}
\begin{equation}\label{oo1}
\Omega_1 = -\frac{\alpha_r}{2}\gamma^2\gamma^4 + \frac{\rho_r}{2}\gamma^1\gamma^3
\end{equation}
\begin{equation}\label{oo2}
\Omega_2 = -\frac{\alpha_\theta}{2}\gamma^2\gamma^4 + \frac{\left(\rho_\theta + e^{-B}\right)}{2}\gamma^1\gamma^3
\end{equation}
\begin{eqnarray}\label{oo3}
\Omega_3 =&& \left(\frac{-\left(e^{-B}-1\right)\cos(-\rho+\theta) + \left(e^{-B}+1\right)\cos(\rho+\theta)}{4}\right)\cosh\alpha\,\gamma^1\gamma^2 \nonumber\\
&&+\left(\frac{\left(e^{-B}-1\right)\cos(-\rho+\theta) - \left(e^{-B}+1\right)\cos(\rho+\theta)}{4}\right)\sinh\alpha\,\gamma^1\gamma^4 \nonumber\\
&&+\left(\frac{-\left(e^{-B}-1\right)\sin(-\rho+\theta) - \left(e^{-B}+1\right)\sin(\rho+\theta)}{4}\right)\cosh\alpha\,\gamma^2\gamma^3 \nonumber\\
&&+\left(-\frac{\left(e^{-B}-1\right)\sin(-\rho+\theta) - \left(e^{-B}+1\right)\sin(\rho+\theta)}{4}\right)\sinh\alpha\,\gamma^3\gamma^4
\end{eqnarray}
\begin{eqnarray}\label{oo4}
\Omega_4 = &&-\frac{A_r e^{(-B+A)}\sin\rho\sinh\alpha}{2}\,\gamma^1\gamma^2
- \frac{A_r e^{(-B+A)}\cos\rho\sinh\alpha}{2}\,\gamma^2\gamma^3 \nonumber\\
&& +\frac{A_r e^{(-B+A)}\sin\rho\cosh\alpha}{2}\,\gamma^1\gamma^4
- \frac{A_r e^{(-B+A)}\cos\rho\cosh\alpha}{2}\,\gamma^3\gamma^4
\end{eqnarray}
\end{subequations}
as those that will be used in the spinorial covariant derivative $D_{j}\psi$ in the Dirac equations. As for the Dirac equations themselves, they will be taken as
\begin{equation}\label{Dirac_equations}
iY^jD_j\psi -m\psi=0
\end{equation}
in which the tetrad fields \eqref{defframe} define the soldering and at the same time encode all information about the background gravitational field. These are the field equations of which we are going to look for solutions next. It is important to specify that our candidate solution will not be a solution of the fully-coupled system of field equations. It will be the exact solution of the Dirac spinorial field equation in a given space-time without backreaction of the spinor on the metric itself. Hence our aim will be focused on the Dirac equations alone, and not on the Einstein equations.
\section{Integrability of a special spinor field}
We set ourselves the task of finding a suitable structure for the spinor field compatible with the above Dirac equations. To guide our intuition, let us have a look at standard solutions of the Dirac equations, that is in the two known integrable cases, given by the Coulomb and elastic potentials \cite{FMB}. Both have the same static spherical symmetry, so our hope is that, albeit different, these two exact solutions might still share some general features that can be taken as basis to constitute a tentative solution also in the case considered in the previous section.

While this hope is justified (and we will in fact find similarities), it is nevertheless very difficult to infer analogous characters between different spinor fields when these spinor fields are compared in different situations. In other words, if we compare spinor fields some of their similar properties might appear different in different frames, and as a consequence we would not see them. Therefore our first step will be that of bringing these two exact solutions in the same frame. This frame will be that in which the spinor field is (locally) at rest and with spin aligned along the third axis. In this circumstance (and only in this circumstance) every difference between the two cases is in fact a real difference. And consequently all analogous characters will be very easily recognized.

When a spinor field is brought into the frame in which it is at rest and with spin aligned along the third axis, it is possible to demonstrate that its general structure is given (in chiral representation) by
\begin{eqnarray}
&\!\psi\!=\!\phi\ e^{i\alpha}\left(\!\begin{tabular}{c}
$e^{\frac{i}{2}b}$\\
$0$\\
$e^{-\frac{i}{2}b}$\\
$0$
\end{tabular}\!\right)
\end{eqnarray}
with $\alpha$ a phase and $\phi$ and $b$ being the module and chiral angle given by $\bar{\psi}\psi\!=\!2\phi^{2}\cos{b}$ and $i\bar{\psi}\gamma^{5}\psi\!=\!2\phi^{2}\sin{b}$ \cite{Fabbri:2020ypd}.

The exact solution for the hydrogen atom is given (for the $1S$ orbital) by a spinor field that, in this frame, has
\begin{eqnarray}
\tan{b}\!=\!-\frac{\alpha}{\Gamma}\cos{\theta}
\end{eqnarray}
with tetrads
\begin{eqnarray}
&e_{0}^{t}\!=\!\Delta\ \ \ \ \ \ \ \ \ \ \ \ \ \ \ \ 
e_{2}^{t}\!=\!\alpha\sin{\theta}\Delta\\
&e_{1}^{r}\!=\!\Gamma\sin{\theta}\Delta\ \ \ \ \ \ \ \ \ \ \ \ \ \ \ \ 
e_{3}^{r}\!=\!\cos{\theta}\Delta\\
&e_{1}^{\theta}\!=\!\frac{1}{r}\cos{\theta}\Delta\ \ \ \ \ \ \ \ \ \ \ \ \ \ \ \ 
e_{3}^{\theta}\!=\!-\frac{\Gamma}{r}\sin{\theta}\Delta\\
&e_{0}^{\varphi}\!=\!\frac{\alpha}{r}\Delta\ \ \ \ \ \ \ \ \ \ \ \ \ \ \ \ 
e_{2}^{\varphi}\!=\!\frac{1}{r\sin{\theta}}\Delta
\end{eqnarray}
in which we have introduced
\begin{equation}
\Delta\!=\![1\!-\!(\alpha\sin{\theta})^{2}]^{-1/2} 
\end{equation}
and $\Gamma\!=\!\sqrt{1-\alpha^{2}}$ as usual \cite{FMB}. The exact solution for the harmonic oscillator is given (in ground state) by a spinor field that, in the very same frame, has
\begin{eqnarray}
\tan{b}\!=\!-\frac{2ar}{a^{2}-r^{2}}\cos{\theta}
\end{eqnarray}
with tetrads
\begin{eqnarray}
&\ \ \ \ \ \ \ \ e_{0}^{t}\!=\!\Delta\ \ \ \ \ \ \ \ \ \ \ \ \ \ \ \ \ \ \ \ \ \ \ \ \ \ \ \ \ \ \ \ e_{2}^{t}\!=\!2ar|r^{2}\!+\!a^{2}|^{-1}\sin{\theta}\Delta\\
&\!\!\!\!\!\!\!\!\!\!\!\!\!\!\!\!e_{1}^{r}\!=\!(a^{2}\!-\!r^{2})|r^{2}\!+\!a^{2}|^{-1}\sin{\theta}\Delta\ \ \ \ \ \ \ \ \ \ \ \ \ \ \ \ \ \ \ \ \ \ \ \ e_{3}^{r}\!=\!\cos{\theta}\Delta\\
&\ \ \ \ \ \ \ \ e_{1}^{\theta}\!=\!\frac{1}{r}\cos{\theta}\Delta\ \ \ \ \ \ \ \ \ \ \ \ \ \ \ \ \ \ \ \ e_{3}^{\theta}\!=\!-\frac{1}{r}(a^{2}\!-\!r^{2})|r^{2}\!+\!a^{2}|^{-1}\sin{\theta}\Delta\\
&\!\!\!\!\!\!\!\!\!\!\!\!\!\!\!\!e_{0}^{\varphi}\!=\!\frac{1}{r}2ar|r^{2}\!+\!a^{2}|^{-1}\Delta\ \ \ \ \ \ \ \ \ \ \ \ \ \ \ \ \ \ \ \ \ \ \ \ \ \ \ \ \ \ \ \ e_{2}^{\varphi}\!=\!\frac{1}{r\sin{\theta}}\Delta
\end{eqnarray}
where now we have defined
\begin{equation}
\Delta\!=\!\left[1\!-\!\left(\frac{2ra}{r^{2}\!+\!a^{2}}\sin{\theta}\right)^{2}\right]^{-1/2}
\end{equation}
and $a\!=\!(E\!-\!m)/2\omega$ with $E^{2}\!=\!m^{2}\!+\!6\omega$ \cite{FMB}. As is clear to a visual comparison, the two types of potentials give rise to two types of solutions that are in fact different, but the spherical symmetry also restricts them to have remarkable analogies. It is indeed obvious that both chiral angles can be written as
\begin{eqnarray}
\tan{b}\!=\!-\frac{1}{X}\cos{\theta}\label{chiral}
\end{eqnarray}
with
\begin{eqnarray}
X_{\mathrm{A}}=\frac{\Gamma}{\alpha}\label{a}
\end{eqnarray}
for the hydrogen atom and
\begin{eqnarray}
X_{\mathrm{O}}=\frac{a^{2}-r^{2}}{2ar}\label{o}
\end{eqnarray}
for the harmonic oscillator. Notice that the same (\ref{a}-\ref{o}) also make the two sets of tetrads coincide. So the hydrogen atom and the harmonic oscillator are two structurally-analogous solutions, the difference being given by the different functional dependence of the function $X$ in the two cases. We can therefore guess that the different form of $X$ is due to the different potentials, but the similar structure of the spinor is due to the spherical symmetry \cite{FC, Fabbri:2021weq}.

Hence, we may extrapolate that such a structure might be a candidate spinor also for other situations of spherical symmetry such as Schwarzschild-like metrics. We thus assume that, in rest-frame and spin-eigenstate, the spinor be
\begin{eqnarray}
\label{scpsie}
\psi=\frac{e^{\frac{\nu}{2}}}{r\sqrt{\sin\theta}}
e^{-i\left(mt-L\varphi\right)}\left(\begin{tabular}{c}
$e^{ib/2}$\\ $0$\\ $e^{-ib/2}$\\ $0$
\end{tabular}\right)
\end{eqnarray}
where the phase has been written in the usual form in terms of the energy $m$ and the angular momentum $L$ for now not better specified. The module has also been written in terms of a function $\nu(r,\theta)$ for easier manipulation of the formulas later on. The chiral angle $b(r,\theta)$ will be chosen according to the above (\ref{chiral}) and so such that
\begin{equation}
\label{beta}
b(r,\theta)=\arcsin\left(\frac{\cos\theta}{\sqrt{\cos^2\theta + \sinh^2\zeta}}\right)
\end{equation}
in terms of a function $\zeta(r,\theta)$ unspecified for the moment. As for the tetrads we assume the functions $\rho$ and $\alpha$ to be
\begin{subequations}
\label{srabll}
\begin{equation}\label{srabll_1}
\rho(r,\theta) = \arcsin\left(\frac{\sin\theta\sinh\zeta}{\sqrt{\cos^2\theta + \sinh^2\zeta}}\right)
\end{equation}
\begin{equation}\label{srabll_2}
\alpha(r,\theta)=-{\rm arcsinh}\left(\frac{\sin\theta}{\sqrt{\cos^2\theta + \sinh^2\zeta}}\right)
\end{equation}
\end{subequations}
in terms of the same $\zeta$ function. This specifies the full structure of the candidate solution.

When the spinor field \eqref{scpsie} is substituted into the Dirac equations \eqref{Dirac_equations} these reduce to the four real equations
\begin{subequations}\label{ped2}
\begin{equation}\label{ped2_1}
\nu_r= e^B \left(2me^{-A} \sin\rho\sinh\alpha - 2m\cos\rho\sin b + \frac{2L\sin\rho\cosh\alpha}{r\sin\theta} - \frac{\rho_\theta}{r} - A_r e^{-B} \right)
\end{equation}
\begin{equation}\label{ped2_2}
\nu_\theta=-2me^{-A}r\cos\rho\sinh\alpha -2mr\sin\rho\sin b - \frac{2L\cos\rho\cosh\alpha}{\sin\theta} + e^{-B}r\rho_r
\end{equation}
\begin{equation}\label{ped2_3}
b_r=e^B \left(2me^{-A}\cos\rho\cosh\alpha - 2m\cos\rho\cos b + \frac{2L\cos\rho\sinh\alpha}{r\sin\theta} - \frac{\alpha_\theta}{r}\right)
\end{equation}
\begin{equation}\label{ped2_4}
b_\theta=2me^{-A}r\sin\rho\cosh\alpha - 2mr\sin\rho\cos b + \frac{2L\sin\rho\sinh\alpha}{\sin\theta} + re^{-B}\alpha_r
\end{equation}
\end{subequations}
and when also the chiral angle \eqref{beta} together with the functions \eqref{srabll} in the tetrads \eqref{defcoframe} and \eqref{defframe} are all plugged in we further obtain that $\zeta_\theta=0$ so that $\zeta=\zeta(r)$ verifying
\begin{equation}\label{exnn_2}
\zeta_r= e^B\left(-2me^{-A}\cosh\zeta + 2m\sinh\zeta + \frac{2L}{r} - \frac{1}{r}\right)
\end{equation}
as well as
\begin{subequations}
\label{exnn}
\begin{equation}\label{exnn_3}
\nu_r= -A_r + \frac{\left(\left(2L - 1\right)\sinh\zeta - 2mr\cos^2\theta\right)\cosh\zeta
- 2me^{-A}r\sinh\zeta\sin^2\theta}{r\left(\cos^2\theta + \sinh^2 \zeta\right)e^{-B}}
\end{equation}
\begin{equation}\label{exnn_4}
\nu_\theta= - \frac{\left(-\left(2L-1\right)\sin^2\theta + 2L\cosh^2\zeta\right)\cos\theta}{\left(\cos^2\theta + \sinh^2\zeta\right)\sin\theta}
\end{equation}
\end{subequations}
for which the conditions of the Schwarz theorem are met, as a direct check would easily show.

It is convenient to express $\zeta(r)$ in terms of another function $Z(r)$ as
\begin{equation}
\label{pzeta}
\zeta(r)=\ln Z(r)
\end{equation}
in terms of which now \eqref{exnn_2}, \eqref{exnn_3} and \eqref{exnn_4} become respectively
\begin{equation}\label{eqzza}
Z_r =e^B\left(m-me^{-A}\right)Z^2 + \frac{e^B}{r}\left(2L-1\right)Z - e^B\left(m+me^{-A}\right)
\end{equation}
\begin{eqnarray}\label{exnn0_3}
\nu_r =&&\frac{\left(-rA_r e^{-B}+2L-1\right)Z^4 + \left(-4mr\sin^2\theta\/e^{-A} - 4mr\cos^2\theta\right)Z^3 -rA_r\left(4\cos^2\theta-2\right)e^{-B}Z^2}{r\left(1+Z^4+\left(4\cos^2\theta-2\right)Z^2\right) e^{-B}} \nonumber\\
&&+ \frac{\left(+4mr\sin^2\theta\/e^{-A}-4mr\cos^2\theta\right)Z -rA_r e^{-B} - 2L+1}{r\left(1+Z^4+\left(4\cos^2\theta-2\right)Z^2\right) e^{-B}}
\end{eqnarray}
\begin{equation}\label{exnn0_4}
\nu_\theta = -\frac{2\cos\theta\left(LZ^4+\left(\left(4L-2\right)\cos^2\theta-2L+2\right)Z^2+L\right)}{\sin\theta\left(1+Z^4+\left(4\cos^2\theta-2\right)Z^2\right)}
\end{equation}
resulting in what does not seem much of a simplification. However, the last two equations can be integrated giving
\begin{equation}\label{solnutg}
\nu= 2\ln N(r) + \frac{1}{2}\ln\left(Z^4 + 2Z^2\cos2\theta+1\right) -2L\ln\left(\sin\theta\right)
\end{equation}
in terms of a function $N(r)$ that has to verify
\begin{equation}\label{fdnurb}
\frac{N_r}{N}= \frac{2mZe^{(B-A)}r -rA_r -2e^B\left(Zmr +L - \frac{1}{2}\right)}{2r}
\end{equation}
yielding the function $N(r)$ once $Z(r)$ is known and with $Z(r)$ being obtained from \eqref{eqzza}. Equation \eqref{eqzza} itself is a Riccati equation which can always be reduced to a linear equation through the Cole--Hopf transformation
\begin{equation}\label{CH_general}
Z(r)= \frac{e^{A(r)-B(r)}}{m-me^{A(r)}}\frac{z'(r)}{z(r)}
\end{equation}
in terms of yet another function $z(r)$ which is the final step of the solutive process. The knowledge of $z(r)$ would entail, via the inverse Cole--Hopf, that of $Z(r)$ which, in turn, gives the $N(r)$ function after integration of \eqref{fdnurb}. Then, functions $\nu(r)$ and $\zeta(r)$ are determined by \eqref{solnutg} and \eqref{pzeta}, and eventually \eqref{beta} and \eqref{srabll} furnish $b(r,\theta)$, $\rho(r,\theta)$ and $\alpha(r,\theta)$ so that the spinor field \eqref{scpsie} and tetrads \eqref{defcoframe} are fully determined. In doing so, the full set of Dirac equations has been effectively reduced to only two equations: one is equation \eqref{fdnurb}, in terms of which we can get $N$ after integration once $Z$ is given; and the other is equation \eqref{eqzza}, giving $Z$ as solution of a differential equation of Riccati type.

As is clear, then, all angular dependence has been separated, leaving only two equations for the radial dependence of the only two functions to be determined. The candidate solution of the form given above, in rest-frame and spin-eigenstate, with a chiral angle \eqref{beta}, together with \eqref{srabll}, has in fact a structure that allows for a complete separation of angular and radial parts, and henceforth for the whole integration of the Dirac equations.

The problem of finding a way to obtain exact solutions for the spinor equations is now solved completely.
\section{Exact solutions in the de Sitter space--time}
In the previous section we have provided a structure for the spinor field that allows the Dirac spinor field equations written in static and spherical metrics to be always separated, and hence integrated at least in general circumstances.

In the present section we will show the general outlines to actually carry on the procedure till the end, finding the explicit solution for one explicit case. As an example of explicit case, we will choose the specific metric given by the de Sitter space--time, for two reasons. One is that the de Sitter metric is sufficiently symmetric to ensure that no unnecessary complications will arise (thus keeping the example at its simplest). The other is that the de Sitter universe is what best approximates the behaviour of cosmological observations soon after the initial singularity and in recent epochs (the two times in which the universe was dominated by a form of cosmological constant).

The de Sitter metric is given by
\begin{equation}\label{DeSitter_metric}
g = \left(1-Kr^2\right)\,dt\otimes dt - \frac{1}{\left(1-Kr^2\right)}\,dr\otimes dr - r^2\,d\theta\otimes d\theta - r^2\sin^2\theta\, d\varphi\otimes d\varphi
\end{equation}
defined in the domain $0\!\leq\!r\!\leq\!1/\sqrt{K}$ for the chosen coordinates. We also take $L=-1/2$ because the spinor in which we are interested is just the spin-$1/2$ spinor after all. Anyway, this choice will be better motivated later.

Now, in this circumstance, equation \eqref{eqzza} assumes the expression
\begin{equation}\label{eqzds}
Z_r =-\frac {mr\left(Z^2+1\right)\sqrt{-kr^2+1}+\left(mrZ^2-mr-2Z\right)\left(Kr^2-1\right)}{r(-Kr^{2}+1)^{3/2}}
\end{equation}
so that after performing the Cole--Hopf transformation
\begin{equation}\label{trchc2e}
Z(r) = \frac{\left(Kr^2-1\right)}{m\left(\sqrt{-Kr^2+1}-1\right)}\frac{z_r(r)}{z(r)}
\end{equation}
we get that \eqref{eqzds} is linearized into
\begin{equation}\label{eqznc}
z_{rr} = -\frac{\left(K^2r^4+Kr^2+2\sqrt{-Kr^2+1}-2\right)z_r}{\left(Kr^2-1\right)\left(Kr^2+\sqrt{-Kr^2+1}-1\right)r} - \frac{Km^2r^2z}{\left(Kr^2-1\right)^2}
\end{equation}
which can always be solved. It is convenient to change the variable according to
\be
\label{posp}
p:=\sqrt{-Kr^2+1}, \qquad p\in[0,1]
\ee
and express the mass of the spinor field as
\be\label{posm}
m=-\alpha\sqrt{K} \qquad {\rm with} \qquad \alpha <0
\ee
for the sake of simplicity. In terms of the new variable and constant, \eqref{eqznc} acquires the form
\be\label{eqzpp}
z_{pp} = \frac{\left[1-p\left(p+1\right)\right]z_p}{p\left(p^2-1\right)} - \frac{\alpha^2z}{p^2}
\ee
which admits the exact solution
\be\label{saz}
z(p) = C_1p^{i\alpha}Q_1(p) + C_2p^{-i\alpha}Q_2(p)
\ee
where
\begin{subequations}
\label{valQ12g}
\begin{equation}
\label{valQ12g_1}
Q_1(p) =\frac{\sqrt{-p+1}\left[\alpha\left(4\alpha^2+1\right)F\left(-\frac{1}{4},\frac{1}{4}+i\alpha;\frac{1}{2}+i\alpha;p^2\right)+
\left(\left(4\alpha^2+\frac{1}{2}\right)+i\alpha\right)p\left(p+1\right)F\left(\frac{3}{4},\frac{5}{4}+i\alpha;\frac{3}{2}+i\alpha;p^2\right)\alpha\right]}{4\alpha^3+\alpha}
\end{equation}
\begin{eqnarray}
\label{valQ12g_2}
Q_2(p) = && -\frac{9\sqrt{-p+1}\left(\frac{1}{2}+i\alpha\right)}{16\alpha^5+40\alpha^3+9\alpha}\left[\frac{1}{9}\left(-\left(2+p\right)+i4\alpha\left(p+1\right)\right)\left(4\alpha^2+9\right)F\left(\frac{3}{4},\frac{1}{4}-i\alpha;\frac{3}{2}-i\alpha;p^2\right)\alpha\right.\nonumber \\
&& \left. - \frac{5}{3}\left(\frac{1}{5}\left(4\alpha^2+\frac{3}{2}\right)-i\alpha\right)p^2\left(p+1\right)F\left(\frac{7}{4},\frac{5}{4}-i\alpha;\frac{5}{2}-i\alpha;p^2\right)\alpha\right]
\end{eqnarray}
\end{subequations}
and with $C_1$ and $C_2$ two complex integration constants. In \eqref{valQ12g}, $F(a,b;c;z)$ is the Gauss hypergeometric function (see for instance the general reference \cite{Stegun}). By expansions to arbitrary orders, we can prove that \eqref{valQ12g_1} and \eqref{valQ12g_2} are complex conjugates of each other. Therefore, the conditions $C_2=\bar{C}_1$ ensures that \eqref{saz} is real as it actually has to be.

The Cole--Hopf transformation \eqref{trchc2e}, in terms of the variable $p$, is
\be
\label{trcho}
Z(p)=\frac{p\sqrt{p+1}}{\alpha\sqrt{-p+1}}\frac{z_p}{z}
\ee
whereas the analogous \eqref{fdnurb}, always in terms of $p$, is
\be\label{eqNuzp}
\frac{N_p}{N} = \frac{z_p}{z} + \frac{-p^2 +2p+1}{2p\left(p^2-1\right)}
\ee
which can be integrated as
\be\label{solNup}
N(p)=\frac{Q\sqrt{K}\sqrt{-p+1}}{\sqrt{p^2+p}}z
\ee
with $Q$ an integration constant. As it stands, the problem is now fully solved, up to variable changes and substitutions.

Once the function $z(p)$ (or $z(r)$) is known, so are the functions $Z(p)$ and $N(p)$ (respectively $Z(r)$ and $N(r)$) from \eqref{trcho} and \eqref{solNup}, and therefore all the remaining functions are determined. Just the same, one must verify that the two integrals given by
\begin{equation}\label{qds}
\int\psi^\dagger\psi\,dV \quad {\rm and} \quad \int\bar\psi\psi\,dV
\end{equation}
converge to ensure the well--posedness of all quantities of physical meaning. In this regard it should be noted that we could have solved the problem by choosing $L$ with an arbitrary value. But at this point, the integrals \eqref{qds} would be seen to converge only for $L<1/2$ strictly. This justifies the choice $L=-1/2$ we made above. The convergence of \eqref{qds} is in turn ensured by that of
\begin{equation}
\label{iien}
I=\int\frac{e^\nu}{r^2\sin\theta}\,dV = 2\pi\int_0^\pi\sin\theta\int_0^{\frac{1}{\sqrt{K}}}N^2(r)\sqrt{1+Z^4(r)+\left(4\cos^2\theta -2\right)Z^2(r)}drd\theta
\end{equation} 
and because
\begin{equation}
\label{iien4}
|I| \leq 4\pi\int_0^{\frac{1}{\sqrt{K}}}N^2(r)\left(Z^2(r)+1\right)dr = \frac{4\pi}{\sqrt K}\int_0^1\frac{N^2(p)\left(Z^2(p)+1\right)p}{\sqrt{-p^2+1}}dp 
\end{equation}
we only need to prove the convergence of the last integral, which is much easier. By employing \eqref{trcho} and \eqref{solNup}, we get the further relation
\begin{equation}
\label{iien5}
|I| \leq \frac{4\pi\sqrt{K}Q^2}{\alpha^2}\int_0^1\frac{p^2\/z_p^2}{\sqrt{-p^2+1}}dp + 4\pi\sqrt{K}Q^2\int_0^1\frac{\sqrt{-p+1}z^2}{\left(p+1\right)^{3/2}}dp
\end{equation}
and in view of \eqref{saz}, we also derive
\begin{equation}
\label{fdexzp}
|z(p)|\leq 2|Q_1(p)||C_1| \quad {\rm and} \quad p\/z_p=\left(i\alpha\/Q_1(p) + p\frac{dQ_1}{dp}\right)p^{i\alpha}C_1 + \left(-i\alpha\/Q_2(p) + p\frac{dQ_2}{dp}\right)p^{-i\alpha}C_2
\end{equation}
which, together with the regularity of the functions $Q_1(p)$ and $Q_{2}(p)$ and their derivatives, prove the convergence of both integrals in \eqref{iien5} in a neighborhood of $p=0$. To discuss instead the convergence in a neighborhood of $p=1$, it is more convenient to perform another change of variable as
\begin{equation}
\label{pv}
p=-v^2 +1 \qquad v\in[0,1]
\end{equation}
leading to
\begin{equation}\label{iien7}
|I| \leq  \frac{4\pi\sqrt{K}Q^2}{\alpha^2}\int_0^1\frac{\left(v^6-4v^4+5v^2-2\right)z_v^2-4z^2(v)v^4\alpha^2}{2\sqrt{v^2\left(2-v^2\right)}v\left(v^2-2\right)}dv
\end{equation}
where now the convergence has to be proved in a neighborhood of $v=0$. By developing $z(v)$ in series at $v=0$ according to
\begin{equation}
z(v)=\sum_nT_nv^n
\end{equation}
one finds that the first three coefficients are given by
\begin{subequations}
\begin{equation}
T_0=\frac{\sqrt{2\pi}\left[C_1(i\alpha+4\alpha^2+1/2)\Gamma^2(i\alpha + 3/2)\Gamma(5/4-i\alpha)
-\pi\bar{C}_1(i\alpha-4\alpha^2-1/2)(\alpha^2+1/4)\Gamma(i\alpha+5/4){\rm sech}(\alpha\pi)\right]}
{(4\alpha^2+1)\Gamma(3/4)\Gamma(5/4-i\alpha)\Gamma(i\alpha+3/2)\Gamma(i\alpha+5/4)}
\end{equation} 
\begin{equation}
\label{solcT2}
T_1=\frac{2i\alpha\sqrt{2}\Gamma(3/4)\left[-C_1\Gamma^2(i\alpha + 1/2)\Gamma(3/4-i\alpha)
+\pi\bar{C}_1\Gamma(i\alpha + 3/4){\rm sech}(\alpha\pi)\right]}
{\Gamma(3/4-i\alpha)\Gamma(i\alpha+1/2)\Gamma(i\alpha+3/4)\sqrt{\pi}}
\end{equation}
\begin{equation}
T_2=0
\end{equation}
\end{subequations}
and the convergence condition is granted by requiring that $T_1=0$ identically. Explicitly this is
\begin{equation}
\label{vxil2} 
f(\alpha)=\frac{{\rm Im}(C_1)}{{\rm Re}(C_1)}=\frac{ad-bc}{ac+bd}
\end{equation}
with
\begin{equation}\label{abcd}
a={\rm Re}(\Gamma(i\alpha+1/2)),\ \ \ \ b={\rm Im}(\Gamma(i\alpha+1/2)),\ \ \ \ 
c={\rm Re}(\Gamma(i\alpha+3/4)),\ \ \ \ d={\rm Im}(\Gamma(i\alpha+3/4))
\end{equation}
and because this condition can always be obtained the convergence is finally established. The solution \eqref{saz}, \eqref{trcho} and \eqref{solNup} is physically acceptable, and as a consequence the found spinor field represents a meaningful material distribution.

We notice two things about the above coefficients. One is that from a purely numerical perspective, we can write relation \eqref{posm} as $\alpha\!\approx\!-l_\mathrm{Universe}/l_\mathrm{Compton}$ where $l_\mathrm{Universe}$ is the radius of the observable Universe and $l_\mathrm{Compton}$ the Compton length of the particle. This means that $\alpha$ is negative and with an absolute value that is impressively large, so that it makes sense to take the limit $\alpha\!\rightarrow\!-\infty$. When this is done \eqref{vxil2} tends to
\begin{equation}
\lim_{\alpha\rightarrow-\infty}f(\alpha)\!=\!1-\sqrt{2}
\end{equation}
which is a remarkably simple value. The second thing is that the vanishing of the coefficient of the quadratic term is granted by the validity of the field equation
\begin{equation}
\label{eqz}
z_{vv}=-\frac{z_{v}v(v^2-3)}{(v+1)(v-1)(v^2-2)}-\frac{4\alpha^2zv^2}{(v-1)^2(v+1)^2}
\end{equation}
in the limit for $v$ that goes to zero. Notice also that with \eqref{eqz} one can also establish the relations tying all other coefficients of the expansion (in this way one can prove for example that $T_{3}=T_{1}/4\!\rightarrow\!0$ and that $T_{4}=-T_{0}\alpha^2/3$).

As a conclusive remark, we point out that the problem would have a similar solution in terms of Gauss hypergeometric functions also in the case of an anti--de Sitter space--time. More in detail, in this case the problem is solved by matching two distinct solutions, one defined in a neighborhood of $r=0$ and the other defined in a neighborhood of $r=\infty$, again ensuring the convergence of \eqref{qds}. The full procedure is perfectly analogous to the one we that have presented here above. We are not going to deepen the discussion on anti--de Sitter metrics, however, since the treatment is more involuted from a purely mathematical perspective without really bringing any new element of clarity.
\section{Conclusion}
In this paper, we have considered the Dirac equation on a background metric of Schwarzschild type, that is static and with spherical symmetry, tackling the problem of finding a special spinor field that could split the Dirac equations in angular and radial parts, so to have them integrated. We have found such a Dirac spinor, given by \eqref{scpsie} with a chiral angle \eqref{beta}, which was defined for a background described by the tetrads \eqref{defframe} with $\rho$ and $\alpha$ given by \eqref{srabll}, and we have proved that the separability was obtained. We also proved that after separation, the full set of equations reduced to only two equations, and we discussed how they could always be integrated, at least in principle.

In practice, we have specialized to the de Sitter universe and, in this case, we have carried on all computations.

The same form (\ref{beta}-\ref{srabll}) can also be used, by construction, to find more general solutions of hydrogen-like atoms.

Quite generally, the procedure we outlined is useful in any case where static/spherical backgrounds are considered.

\

{\bf Funding information and acknowledgements}. This work has been carried out in the framework of activities of the INFN Research Project QGSKY. The work of L.F. is funded by Next Generation EU through the project ``Geometrical and Topological effects on Quantum Matter (GeTOnQuaM)''.

\

{\bf Authors individual contributions}. All authors have contributed equally to the computations and to the writing.

\

{\bf Conflict of interest}. The authors declare no conflict of interest.

\

{\bf Data availability}. This manuscript has no data available.


\begin{thebibliography}{50}
\bibitem{hasan2010colloquium} 
M. Z. Hasan, C. L. Kane, ``Topological insulators'', 
\textit{Rev.Mod.Phys.} \textbf{82}, 3045 (2010).
\bibitem{boada2014dirac}
O. Boada, A. Celi, M. Lewenstein, J. Rodr\'\i{}guez-Laguna, 
J. I. Latorre, ``Quantum simulation\\ of non-trivial 
topology'', \textit{New J. Phys.} \textbf{17}, 045007 (2015).
\bibitem{suchet2016analog}
D. Suchet, M. Rabinovic, T. Reimann, N. Kretschmar, 
F. Sievers, C. Salomon, J. Lau, O. Goulko, C. Lobo, 
F. Chevy,\\ ``Analog simulation of Weyl particles with 
cold atoms'', \textit{EPL} \textbf{114}, 26005 (2016).
\bibitem{novoselov2005two}
K.S.Novoselov, A.K.Geim, S.V.Morozov, D.Jiang, 
M.I.Katsnelson, I.V.Grigorieva, S.V.Dubonos, 
A.A.Firsov,\\ ``Two-dimensional gas of massless Dirac 
fermions in graphene'', \textit{Nature}, \textbf{438}, 197 (2005).
\bibitem{katsnelson2006chiral}
M.I.Katsnelson, K.S.Novoselov, A.K.Geim, ``Chiral 
tunnelling and the Klein paradox\\ in graphene'', 
\textit{Nat. Physics} \textbf{2}, 620 (2006).
\bibitem{schwerdtfeger2015relativistic}
P. Schwerdtfeger, L. F. Pa\v{s}teka, A. Punnett, 
P.O.Bowman, ``Relativistic and quantum \\
electrodynamic effects in superheavy 
elements'', \textit{Nucl. Phys. A} \textbf{944}, 551 (2015).
\bibitem{pavsteka2017relativistic}
L. F. Pa\v{s}teka, E. Eliav, A. Borschevsky, U. Kaldor, 
P. Schwerdtfeger, ``Relativistic Coupled Cluster 
Calculations with\\ Variational Quantum 
Electrodynamics Resolve the Discrepancy between 
Experiment and Theory Concerning the\\ Electron 
Affinity and Ionization Potential of Gold'', 
\textit{Phys. Rev. Lett.} \textbf{118}, 023002 (2017).
\bibitem{autschbach2012perspective}
J. Autschbach, ``Perspective: relativistic effects'', 
\textit{J. Chem. Phys.} \textbf{136}, 150902 (2012).
\bibitem{Cianci:2019oor}
Cianci, R., Fabbri, L., Vignolo, S., ``Axially-Symmetric Exact Solutions for\\
Flagpole Fermions with Gravity'', \textit{Eur. Phys. J. Plus} \textbf{135}, 131 (2020).
\bibitem{Cianci:2016pvd}
Cianci, R., Fabbri, L., Vignolo, S., ``Critical exact solutions for self-gravitating\\
Dirac fields'', \textit{Eur. Phys. J. C} \textbf{76}, 595 (2016).
\bibitem{Cianci:2015pba}
Cianci, R., Fabbri, L., Vignolo, S., ``Exact solutions for Weyl fermions\\
with gravity'', \textit{Eur. Phys. J. C} \textbf{75}, 478 (2015).
\bibitem{Soler:1970xp}
Soler, M., ``Classical, stable, nonlinear spinor field with positive rest energy'', 
\textit{Phys. Rev. D} \textbf{1}, 2766 (1970).
\bibitem{Ranada:1983zqr}
Ranada, A. F., ``Classical nonlinear Dirac field models of extended particles'', 
\textit{Math. Phys. Stud.} \textbf{4}, 271 (1983).
\bibitem{Saha:2018ufp}
Saha, B., ``Spinor fields in spherically symmetric space-time'', 
\textit{Eur. Phys. J. Plus} \textbf{133}, 461 (2018).
\bibitem{Bronnikov:2019nqa}
K. A. Bronnikov, Y. P. Rybakov, B. Saha, ``Spinor fields in spherical symmetry: Einstein-Dirac\\ and other space-times'', \textit{Eur. Phys. J. Plus} \textbf{135}, 124 (2020).
\bibitem{Herdeiro:2017fhv}
Herdeiro, C. A. R., Pombo, A. M., Radu, E., ``Asymptotically flat scalar,\\
Dirac and Proca stars: discrete vs. continuous families of solutions'', 
\textit{Phys. Lett. B} \textbf{773}, 654 (2017).
\bibitem{Herdeiro:2019mbz}
Herdeiro, C., Perapechka, I., Radu, E., Shnir, Ya., ``Asymptotically flat spinning scalar,\\
Dirac and Proca stars'', \textit{Phys. Lett. B} \textbf{797}, 134845 (2019).
\bibitem{Dzhunushaliev:2019kiy}
Dzhunushaliev, V., Folomeev, V., ``Dirac star in the presence of Maxwell and Proca fields'', 
\textit{Phys. Rev. D} \textbf{99}, 104066 (2019).
\bibitem{Blazquez-Salcedo:2019qrz}
Bl\'{a}zquez-Salcedo, J. L., Knoll, C., Radu, E., ``Boson and Dirac stars in $D\geq 4$ dimensions'', \textit{Phys. Lett. B} \textbf{793}, 161 (2019).
\bibitem{bs-k}
Bl\'{a}zquez-Salcedo, J.L., Knoll, C. ``Constructing spherically symmetric Einstein-Dirac systems with\\ multiple spinors: Ansatz, wormholes and other analytical solutions'',
\textit{Eur. Phys. J. C} \textbf{80}, 174 (2020).
\bibitem{fss}
Finster, F., Smoller, J., Yau, S., ``Particle-like solutions of the Einstein-Dirac equations'', \textit{Phys. Rev. D} \textbf{59}, 104020 (1999).
\bibitem{fss1}
F. Finster, J. Smoller, S. Yau, ``Non-existence of black hole solutions for a 
spherically symmetric,\\ static Einstein-Dirac-Maxwell system'',
\textit{Commun. Math. Phys.} \textbf{205}, 249 (1999).    
\bibitem{fss2}
F. Finster, J. Smoller, S. Yau, ``Non-existence of time periodic solutions of the Dirac equation\\
in a Reissner-Nordstrom black hole background'', \textit{J. Math. Phys.} \textbf{41}, 2173 (2000).
\bibitem{fss3}
F. Finster, J. Smoller, S. Yau, ``Absence of stationary, spherically symmetric black hole
solutions for\\ Einstein-Dirac-Yang-Mills equations with angular momentum'',
\textit{Adv. Theor. Math. Phys.} \textbf{4}, 1231 (2002).
\bibitem{Bagchi:2023cpz}
B. Bagchi, A. Gallerati, R. Ghosh, ``Dirac equation in curved spacetime: the role of\\
local Fermi velocity'', \textit{Eur. Phys. J. Plus} \textbf{138}, 1037 (2023).
\bibitem{Dirac}
P. A. M. Dirac, ``The Electron Wave Equation in De-Sitter Space'', \textit{Annals of Mathematics} \textbf{36}, 657 (1935).
\bibitem{H}
K. C. Hannabuss, ``The Dirac equation in de Sitter space'', \textit{J. Phys. A: Gen. Phys.} \textbf{2}, 274 (1969).
\bibitem{KT}
V. V. Klishevich, V. A. Tyumentsev, ``On the solution of the Dirac equation in\\
de Sitter space'', \textit{Class. Quant. Grav.} \textbf{22}, 4263 (2005).
\bibitem{Cotaescu:2007xv}
I. Cotaescu, ``Dirac fermions in de Sitter and anti-de Sitter backgrounds'',
\textit{Rom. J. Phys.} \textbf{52}, 895 (2007).
\bibitem{Kanno:2016qcc}
S. Kanno, M. Sasaki, T. Tanaka, ``Vacuum State of the Dirac Field in de Sitter Space\\
and Entanglement Entropy'', \textit{JHEP} \textbf{03}, 068 (2017).
\bibitem{Dappiaggi:2019oew}
C. Dappiaggi, F. Finster, S. Murro, E. Radici, ``The fermionic signature operator in\\
de Sitter spacetime'', \textit{J. Math. Anal. Appl.} \textbf{485}, 123808 (2020).
\bibitem{FMB}
L. Fabbri, F. Moulin, A. Barrau, ``Non-trivial effects of sourceless forces for spinors: toward\\
an Aharonov-Bohm gravitational effect?'', \textit{Eur.Phys.J.C}\textbf{79}, 875 (2019).
\bibitem{Fabbri:2020ypd}
Luca Fabbri, ``Spinors in Polar Form'', \textit{Eur. Phys. J. Plus} \textbf{136}, 354 (2021).
\bibitem{FC}
Luca Fabbri, A. G. Campos, ``Angular-radial integrability of Coulomb-like\\
potentials in Dirac equations'', \textit{J. Math. Phys.}\textbf{62}, 113505 (2021).
\bibitem{Fabbri:2021weq}
L.Fabbri, R.Cianci, S.Vignolo, ``A square-integrable spinor solution to\\
non-interacting Dirac equations'', \textit{AIP Adv.}\textbf{11}, 115314 (2021).
\bibitem{Stegun}
M. Abramowitz, I. A. Stegun, {\it Handbook of Mathematical Functions:\\
with Formulas, Graphs, and Mathematical Tables} (Dover Publications Inc., 1965).
\end{thebibliography}
\end{document}